\documentclass[%
 pra,reprint,
 superscriptaddress, 
  amsmath,amssymb,
 aps,
]{revtex4-2}

\usepackage{physics} 

\usepackage{comment}
\usepackage{xcolor}

\usepackage{graphicx}
\usepackage{dcolumn}

\usepackage{hyperref}
\usepackage{orcidlink} 
\hypersetup{
    colorlinks,
    linkcolor={blue!80!black},
    citecolor={blue!50!black},
    urlcolor={blue!80!black}
}

\begin{document}
\title{Dark state spectroscopy in nonlinear waveguide quantum electrodynamics
}

\author{Shay Nadel\,\orcidlink{0009-0008-6761-5356}}
\affiliation{Andrew and Erna Viterbi Department of Electrical \& Computer Engineering, Technion - Israel Institute of Technology, Haifa 32000, Israel}
\affiliation{Helen Diller Quantum Center, Technion - Israel Institute of Technology, Haifa 32000, Israel}

\author{Amir Sivan\,\orcidlink{0000-0002-0498-8867}}
\affiliation{Andrew and Erna Viterbi Department of Electrical \& Computer Engineering, Technion - Israel Institute of Technology, Haifa 32000, Israel}
\affiliation{Helen Diller Quantum Center, Technion - Israel Institute of Technology, Haifa 32000, Israel}

\author{Aviv Karnieli\,\orcidlink{0000-0002-4056-4455}}
\email{karnieli@technion.ac.il} 
\affiliation{Andrew and Erna Viterbi Department of Electrical \& Computer Engineering, Technion - Israel Institute of Technology, Haifa 32000, Israel}
\affiliation{Helen Diller Quantum Center, Technion - Israel Institute of Technology, Haifa 32000, Israel}

\date{\today}

\preprint{APS/123-QED}

\begin{abstract}
Quantum systems face a fundamental trade-off:~they must remain decoupled from the environment to maintain long coherence times, yet they require interactions with the environment to be accessible for measurement.~As a prime example, emitter arrays coupled to waveguides facilitate collective modes that, owing to interference, can suppress radiation into the waveguide. While complete destructive interference creates perfectly dark states with infinite lifetimes, their inherent decoupling makes them unmeasurable in standard waveguide quantum electrodynamics. Consequently, current approaches must rely on system non-idealities that permit measurement but limit the coherence times. In this work, we lift this limitation by proposing the use of weakly squeezed light generated in $\chi^{(2)}$ nonlinear waveguides for the spectroscopy of completely dark states. We show that the fluorescence spectrum probes transitions between the dressed dark states of the emitter array. This work paves the way towards the measurement and control of dark states, with applications for robust quantum memories, computation, and communication.
\end{abstract}

\maketitle

\section{\label{sec:Introduction}Introduction}

Quantum technologies, such as quantum computing, memory, and sensing, rely profoundly on the generation and preservation of entanglement across many qubits. These many-body entangled systems face a fundamental trade-off: they must remain decoupled from the environment to maintain long coherence times, yet they require sufficient interaction with the environment to remain accessible for measurement and control \cite{zurek2003decoherence,haroche2013nobel,divincenzo2000physical,clerk2010introduction,georgescu2014quantum}.

Waveguide quantum electrodynamics (WQED) has emerged as a prime paradigm to navigate this tension \cite{shermet2023waveguidereview,chang2018colloquium,blais2021circuit,roy2017colloquium,gu2017microwave,kockum2018decoherence}. In these systems, arrays of quantum emitters coupled to a one-dimensional photonic waveguide exhibit collective behaviors governed by photon-mediated dipole-dipole interactions \cite{asenjo2017exponential}.
Depending on the spacing between the emitters, their orientation, and their electromagnetic environment, constructive and destructive interference of individual emissions manifests as superradiant and subradiant collective states, respectively \cite{shermet2023waveguidereview,lalumiere2013inputoutput,vanloo2013longrange,dicke1954coherence, gross1982superradiance,pichler2015quantum, sivan2019enhanced, john1995localization, pustovit2010plasmon}.
For an ideal system with exact wavelength spacing, perfectly destructive interference yields completely dark states with infinite lifetimes. 

However, this absolute decoupling from the environment poses a severe challenge: completely dark states are inherently inaccessible to standard optical probes.
Consequently, the detection of subradiant states in standard linear waveguides has relied profoundly on system non-idealities that break the perfect destructive interference \cite{lalumiere2013inputoutput,vanloo2013longrange}, indirect measurements via local drives \cite{zanner2022coherentcontrol}, or probe qubits 
\cite{mirhosseini2019cavity}. For example, transmission spectroscopy can detect subradiant states as narrow resonance features, but a completely dark state possesses an infinitely narrow linewidth, rendering it invisible to a probe \cite{vanloo2013longrange}. Similarly, incoherent scattering (or fluorescence) measurements inherently rely on the state not being completely dark, requiring some degree of radiative leakage \cite{lalumiere2013inputoutput,albrecht2019subradiant,vanloo2013longrange} or non-ideal emitter spacings \cite{poshakinskiy2021dimerization,fang2015twophotonscattering}. While recent advances have enabled the measurement of transitions between subradiant states via two-time correlation functions \cite{albrecht2019subradiant} or local coherent control \cite{zanner2022coherentcontrol}, completely dark states remain fundamentally hidden in linear setups.

A path to overcoming this limitation is the addition of optical nonlinearity within the waveguide \cite{karnieli2025,wang2024long,wang2026amplifying,saravi2017atom}. By weakly pumping a nonlinear waveguide medium at twice the natural frequency of the emitters, the pump light is converted into squeezed waveguide modes that couple to the emitters. This squeezed light mediates effective, long-range  interactions even for emitters spaced by a multiple of a wavelength, where coherent interactions mediated by linear waveguides typically wash out.
For weak pump amplitudes, bright states can be adiabatically eliminated, and the system dynamics are restricted to a decoherence-free subspace populated entirely by dark-state excitations \cite{karnieli2025}. 

In this work, we demonstrate that a $\chi^{(2)}$ nonlinear waveguide acting as a traveling-wave parametric amplifier, enables the spectroscopic measurement of completely dark states and the transitions between them.
When the nonlinear waveguide is pumped, it supports a parametric gain that breaks the permutation symmetry between emitters.
In this way, the otherwise perfectly dark states can now couple to the squeezed waveguide mode.
At steady state, the squeezed vacuum enables transitions between the dressed dark eigenstates of the interaction Hamiltonian. This results in emitted photons carrying distinct signatures of the dark manifold.
Using input-output theory, we calculate the incoherent spectrum, and show that it unveils the transition frequencies between dressed dark eigenstates in both ideal and realistic experimental systems. Our work paves the way towards coherent control and readout of dark states in WQED.
Accessing this decoherence-free manifold of completely dark states has implications in quantum memories and computation \cite{asenjo2017exponential,manzoni2018optimization,gonzalez2015deterministic,Lidar1998DecoherenceFree,paulisch2016universal,tabares2023variational}, precise metrology \cite{zafra2025subradiant,shi2025quantum}, and pushes the limits of subradiance-protected communication \cite{needham2019subradiance}.

\section{\label{sec:Model}Model}

\begin{figure}
\includegraphics[width=\linewidth]{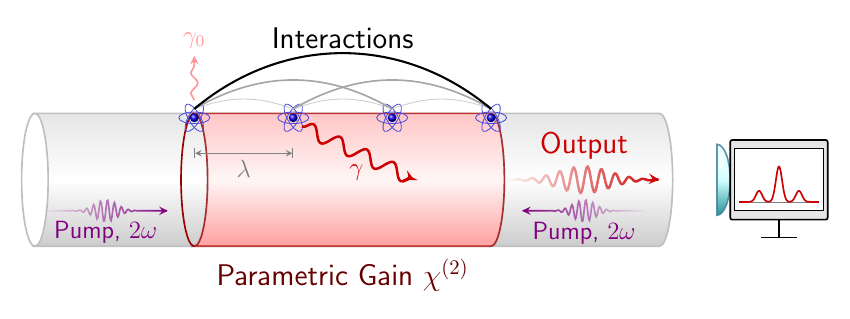}
\caption{\label{fig:System Illustration}
\textbf{System illustration.}
A waveguide with a nonlinear section of length $L$ (red cylinder), pumped from the left and the right with coherent light at frequency $2\omega$ (violet).
The $\chi^{(2)}$ nonlinearity introduces a parametric gain, generating two photons in frequency $\omega$ and wavelength $\lambda$ within the nonlinear segment.
Identical two-level emitters with a transition energy of $\hbar \omega$ are spaced by a wavelength and coupled to the nonlinear segment (depicted is the case of 4 emitters in blue). 
The output signal is collected from the right side of the waveguide.
The emitter array can radiate with rate $\gamma$ into the waveguide modes, and with rate $\gamma_0$ to free-space. 
Within the nonlinear section, owing to the parametric gain, photons mediate interactions between the emitters that increase with distance, illustrated as thicker lines for stronger interactions (black).
}
\end{figure}

We consider an infinite one-dimensional waveguide containing a $\chi^{(2)}$ nonlinear section of length $L$. 
The nonlinear section is driven by two counter-propagating coherent pumps at frequency $2\omega$, and acts as a traveling-wave parametric amplifier.
Within the nonlinear section, each pump is downconverted into two waveguide modes with frequency $\omega$ and wavelength $\lambda$.
$N$ identical two-level-systems with transition energy of $\hbar \omega$ are coupled to the nonlinear section.
These emitters are positioned along the nonlinear section at positions $z_j = \lambda \left( j-1 \right)$, where $j={1,...,N}$. They are spaced by $\lambda$ and for simplicity, the first and last emitters are located at the edges of the nonlinear section, imposing $L = \left( N-1 \right) \lambda$.
A system illustration is presented in Fig.~\ref{fig:System Illustration}. Our goal is to calculate the spectrum of the light emitted by this system, which carries the signature of dark states, as will be explained below. 


To determine how this parametric drive allows for the readout of dark states, we structure our analysis in two steps. First, we derive the input-output relations governing the propagation of the photonic field through the nonlinear medium containing the emitter array.
This allows us to connect the temporal correlations of light at the output with correlations of the emitters. 
From these correlations, we will calculate the emitted spectrum. 
Second, we calculate the emitter correlations. 
We use the SLH formalism to derive the interaction Hamiltonian and jump operators that govern the emitter array dynamics~\cite{combes2017slh,carmichael1999statistical}.

The full Hamiltonian in the interaction picture is the nonlinear WQED Hamiltonian \cite{karnieli2025}:
\begin{equation}
\begin{split}
    &H_{\mathrm{WQED}} 
    =
    \hbar \sum_{s\in\rightleftarrows} 
    \bigg[
        \int dza_{s}^{\dagger} 
        \left( z,t \right) 
        \left( 
            -d_{s}v_{g} 
            \frac{\partial}{\partial z} 
        \right)
        a_{s} \left(z,t\right)
    \\
    & \quad
    +
    \intop_{0}^{L} dz
    \kappa
    e^{id_{s} \Delta k\left(z-L/2\right)} 
    a_{s}^{2} 
    \left( z,t \right)
    +
    \mathrm{h.c.}
    \\
    & \quad
    +
    \sqrt{v_{g} \gamma}
    \sum_{j=1}^{N} 
    e^{i\phi_s} e^{ id_{s} kz_{j} }
    a_{s} \left( z_{j},t \right) 
    \sigma_{j}^{\dagger} 
    +
    \mathrm{h.c.}
    \bigg]
    .
\end{split}
\label{eq:original_hamiltonian}
\end{equation} 
where $a_s$ are the right- and left-moving photonic field operators satisfying the bosonic canonical commutation relations
$
    \left[ 
        a_s \left( z,t \right)
        ,
        a_{s^\prime}^\dagger \left( z^\prime,t \right)
    \right]
    =
    \delta \left( z-z^\prime \right)
    \delta_{s,s^\prime}
$.
The group velocity within the waveguide is
$v_g$, and $k= 2\pi / \lambda$ is the wave number.
$d_{\rightarrow}= 1,d_{\leftarrow}= -1$ are the direction coefficients, and $\kappa$ denotes the nonlinear coupling parameter, which is proportional to $\chi^{(2)}$ and the strength of the pump fields at $2\omega$.
$\Delta k$ is the phase mismatch, which will be considered to be zero in this work.
Here, $\sigma_j$ is the spin annihilation operator acting on emitter $j$, and $\gamma$ is the decay rate into the waveguide.
For a nonlinear section commensurate with the resonance wavelength, the relative phases $\phi_{\rightarrow} =0,\phi_{\leftarrow}=kL$ are identically zero modulo $2\pi$.
$\gamma$ is the single-emitter decay rate into the waveguide.
In Sec.~\ref{sec:Experimental considerations}, we further consider the case of a nonideal system that allows the decays of emitters to free space with decay rate $\gamma_0 \ll \gamma$~\cite{blais2021circuit,hoi2011demonstration,astafiev2010resonance}.

\textit{Photonic correlations.} 
We start by using the Heisenberg equations of motion 
$
    i \hbar \frac{d}{dt}a_{s} 
    = 
    \left[ 
    a_{s}, H_{\mathrm{WQED}}
    \right]
$
to propagate the photonic field operators along the nonlinear section.
We define the right-moving input and output fields as 
$
    a_{\rightarrow,\mathrm{in}} \left( t \right)
    =
    a_{\rightarrow} \left( 0,t \right)
$
and
$
    a_{\rightarrow,\mathrm{out}} \left( t \right)
    =
    a_{\rightarrow} \left( L,t \right)
$. 
Using the total squeezing parameter 
$
r = 2 |\kappa | L/v_g
$,
the gain functions 
$
p_{\rightarrow} \left( z \right)
=
\cosh \left(  rz/L \right)
$,
$
q_{\rightarrow} \left( z \right)
=
i \kappa / | \kappa | \sinh \left(  rz/L \right)$,
the Bogoliubov-transformed system operators 
$
    c_{j,\rightarrow} \left( t \right )
    =
    p_{\rightarrow} \left(L- z_j \right) e^{-ikz_j} \sigma_j \left( t \right)
    +
    q^*_{\rightarrow} \left(L- z_j \right) e^{ikz_j} \sigma_j^\dagger \left( t \right)
$,
and by neglecting retardation between emitters (assuming the retardation time to be much smaller than the decay time of the emitters $v_g/L \gg \gamma$~\cite{caneva2015quantum}), we obtain the input-output relations:
\begin{equation}
\begin{split}
a_{\rightarrow,\mathrm{out}} \left( t \right)
&=
p_{\rightarrow} \left( L \right) 
a_{\rightarrow, \mathrm{in}} \left( t \right)
+
q_{\rightarrow}^* \left( L \right) 
a_{\rightarrow, \mathrm{in}}^\dagger \left( t \right)
\\&-
i \sqrt{\frac{\gamma}{v_g}}\sum_j c_{\rightarrow,j} \left( t \right)    
\end{split}
\end{equation}
as discussed in the Supplementary Material (SM) Sec.~S2~\cite{supp}.
Similarly, the left-moving field relations are given by the reflection of the right-moving relations.
Following the derivation done in the SM Sec.~S1~\cite{supp}, the output photon correlations in steady state are
\begin{equation}
\begin{split}
    \left< 
        a_{s,\mathrm{out}}^\dagger \left( \tau \right)
        a_{s,\mathrm{out}} \left( 0 \right)
    \right>
    &= 
    \frac{\left| q_s \left( L \right) \right|^2}{v_g} \delta \left( \tau \right) 
    + 
    \frac{\gamma}{v_g} 
    \left< 
        \mathcal{C}_{s}^\dagger \left( \tau \right) 
        \mathcal{C}_{s} \left( 0 \right) 
    \right>
     \\
    &+ 
    \left| q_s \left( L \right) \right|^2
    \frac{\gamma}{v_g} 
    \left< 
        \left[ 
            \mathcal{C}_{s}^\dagger \left( \tau \right)
            ,
            \mathcal{C}_{s} \left( 0 \right) 
        \right]
    \right> 
     \\
    &+
    q_{s}^* \left( L \right)
    p_{s} \left( L \right)
    \frac{\gamma}{v_g}
    \left< 
        \left[ 
            \mathcal{C}_{s}^\dagger \left( \tau \right)
            ,
            \mathcal{C}_{s}^\dagger \left( 0 \right) 
        \right]
    \right> 
\end{split}
\label{eq:Photon-correlations}
\end{equation} 
where 
$
\mathcal{C}_s 
=
\sum_j c_{s,j}
$.
The first term is a white noise associated with the infinite-bandwidth squeezed vacuum generated in the background.
In the spectral domain, this term will contribute a constant noise background in the output spectrum, irrespective of the light radiated by the emitters, and we therefore proceed to drop it henceforth.
To calculate the incoherent power spectrum of the photons exiting the waveguide to the right at steady-state, we take the Fourier transform of the correlations~\cite{gardiner1985input,carmichael1999statistical} 
\begin{equation}
    S 
    \left(
        \omega
    \right)
    =
    \int d\tau 
    e^{-i\omega \tau}
    \left< 
        a_{\rightarrow,\mathrm{out}}^\dagger \left( \tau \right)
        a_{\rightarrow,\mathrm{out}} \left( 0 \right)
    \right>.  
    \label{eq:power spectrum}
\end{equation}
We evaluate this expression by applying the quantum regression theorem \cite{lax1963quantumregressiontheorem} to the system operator correlations, after substituting Eq.~(\ref{eq:Photon-correlations}).

\textit{System dynamics.}
We now wish to find the open quantum system dynamics for the emitters, generally given by the Lindblad master equation
\begin{equation}
    \dot{\rho}=-\frac{i}{\hbar} [H_{\mathrm{int}},\rho] +\sum_{s\in \rightleftarrows}L_s\rho L_s^{\dagger} -\frac{1}{2}\lbrace L_s^{\dagger}L_s,\rho\rbrace
    \label{eq:master}
\end{equation}
where $H_{\mathrm{int}}$ is the interaction Hamiltonian between the emitters, $\rho$ is the density matrix of the emitters, and $L_s$ are the jump operators \cite{plenio1998quantum}. 

To find the master equation, we apply the Born-Markov and rotating-wave approximations to trace out the propagating photonic modes.
We do so by using the SLH formalism \cite{combes2017slh}, and obtain the interaction Hamiltonian and jump operators (See SM Sec.~S3~\cite{supp}):
\begin{subequations} \label{eq:SLH_dynamics}
\begin{align}
    H_{\mathrm{int}} 
    &=
    \frac{\hbar\gamma}{2} \sum_{i,j} 
    \sinh \left( r \frac{|i-j|}{N-1} \right) 
    \left( 
        \sigma_i \sigma_j 
        +
        \sigma_i^\dagger \sigma_j^\dagger 
    \right), 
    \label{eq:SLH Hamiltonian} \\
    L_{s} 
    &=
    \sqrt{\gamma} 
    \left[ 
    p_{s} \left( z_j\right) e^{-id_s kz_j}
    \sigma_j
    +
    q_{s}^* \left( z_j\right)e^{id_sk z_j} \sigma_j^\dagger
    \right]
    \label{eq:SLH Jump operator}
\end{align}
\end{subequations}
The interaction Hamiltonian $H_{\mathrm{int}}$ displays two striking features.
First, the interactions between the emitters increase with distance, as illustrated in Fig.~\ref{fig:System Illustration}, owing to the parametric gain that the photons experience as they travel through the waveguide.
Second, the Hamiltonian consists of interactions that create and annihilate excitation pairs, as opposed to the spin-exchange interaction common in WQED~\cite{shermet2023waveguidereview}.



In Sec.~\ref{sec:results} we use Eqs.~(\ref{eq:Photon-correlations}-\ref{eq:SLH_dynamics}) to numerically calculate the steady-state incoherent spectrum.
The jump operators (Eq.~(\ref{eq:SLH Jump operator})) permit transitions between the emitter eigenstates of the interaction Hamiltonian (Eq.~(\ref{eq:SLH Hamiltonian})), allowing photons to be emitted into the waveguide. 
We show that the photons in the waveguide encode information about the system, enabling the detection of transitions between dark states.

\section{\label{sec:results}Results}

\begin{figure}
\includegraphics[width=\linewidth]{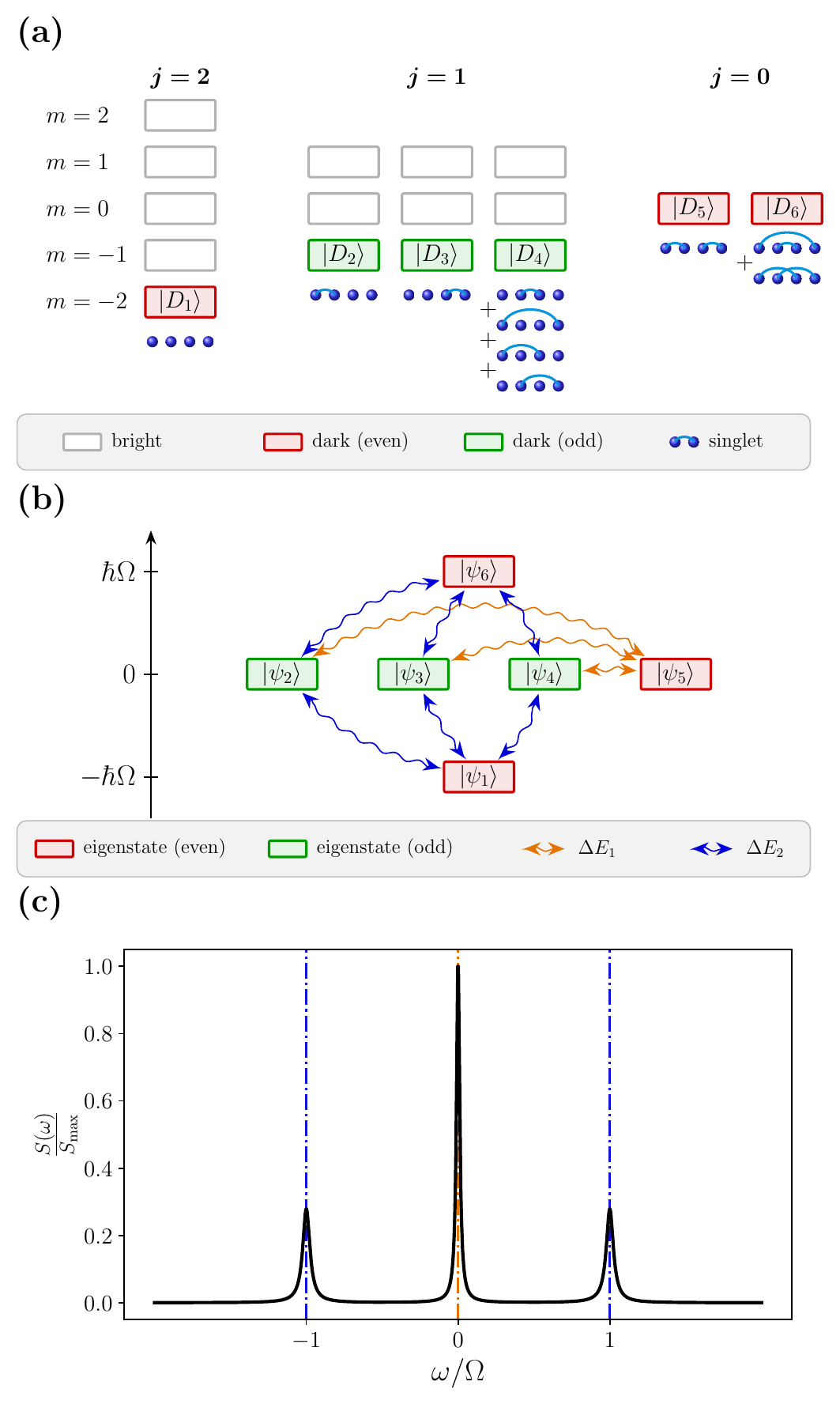}
\caption{\label{fig:DressedDiagram}\textbf{Dark-state manifold and steady-state spectroscopy for an ideal system ($\gamma_0=0$) and $N=4$ emitters.}
\textbf{(a)} Classification of the collective modes within the total angular momentum basis $\ket{j,m}$.
The dark states $\ket{D_i}$, which are the $\ket{j,-j}$ states, are color-coded depending on the value of their $j$-parity.
\textbf{(b)} Dark eigenstates $\ket{\psi_i}$ of the Hamiltonian $H_{\mathrm{int}}$ (Eq.~(\ref{eq:SLH Hamiltonian})) projected onto the dark-state manifold are illustrated with respect to their eigenvalue relative to the emitter transition.
Only transitions between different $j$-parity values are allowed by the jump operators (Eq.~(\ref{eq:SLH Jump operator})).
The allowed transitions are depicted as curly lines, with two distinct colors matching $\Delta E_1 =0, \Delta E_2 =\hbar \Omega$.
\textbf{(c)} Black curve is the numerically calculated incoherent power spectrum of the right-moving waveguide mode at steady state, for a weak squeezing parameter $r=0.01$ and decay rate set to $\gamma=1.0$. 
The peaks appear at $\omega = \Omega\approx 0.385 \gamma r$.
Orange dashed line corresponds to $\Delta E_1/\hbar = 0$, and blue dashed lines corresponds to $\pm \Delta E_2/\hbar=\pm \Omega$, matching the transitions in panel (b).
}
\end{figure}

Our main result is that weakly squeezed light can be used to measure the otherwise completely inaccessible dark states, as the transition frequencies between dark states appear in the incoherent steady-state power spectrum.

Before we dive into the results, we first overview the properties of the linear system (i.e., with the pumps turned off in Fig.~\ref{fig:System Illustration}).
When the emitters are spaced by a multiple of the wavelength, one can classify the many-body states of the system into two categories: Bright states and dark states. States that can collectively radiate into the waveguide are called bright states, while states that cannot collectively radiate into the waveguide are referred to as dark states.
Such dark states are annihilated by the total angular momentum lowering operator $S_- = \hbar /2 \sum_j \sigma_j$, and are therefore always the lowest-lying eigenstates in the angular momentum ladder. 
Fig.~\ref{fig:DressedDiagram}(a) depicts the collective modes for four emitters in the basis of total angular momentum $\ket{j,m}$, with the dark states being the $\ket{j,-j}$ states at each $j$ section. In the context of Eq.~(\ref{eq:master}), the interaction Hamiltonian between the emitters vanishes at wavelength spacing, and the jump operators coincide with $S_-$~\cite{lalumiere2013inputoutput}, rendering the dark states inaccessible to probing using scattered or emitted light.

This dynamical picture changes for the nonlinear system (with the pumps turned on in Fig.~\ref{fig:System Illustration}). 
The new interaction Hamiltonian of Eq.~(\ref{eq:SLH Hamiltonian}) contains the annihilation and creation of emitter excitation pairs. 
The new jump operators of Eq.~(\ref{eq:SLH Jump operator}), mixing $\sigma_j$ and $\sigma_j^\dagger$, break the permutation symmetry of the system and permit the originally dark states to couple to the waveguide modes. 
Moreover, the steady-state of the system is unique, and for small squeezing ($r\ll 1$), consists almost exclusively of dark states (See SM Sec.~S4~\cite{supp}).
The physical origin of the exclusively dark steady state lies in the relatively slow dynamics of the dark states, whose rate of change scales with $\gamma r$ \cite{karnieli2025}. 
The bright states, on the other hand, decay back to the dark subspace at a much faster rate, and can be adiabatically eliminated, as was shown in Ref.~\cite{karnieli2025}.



Consequently, we project the interaction Hamiltonian (Eq.~(\ref{eq:SLH Hamiltonian})) onto the dark-state subspace $PH_{\mathrm{int}}P$, where $P$ is the projector onto the dark subspace. 
As the interaction Hamiltonian commutes with the $j$-parity operator 
 $
    \Pi 
    =
    \bigotimes_{i} \sigma_i^z 
$, we define the dark eigenstates $\ket{\psi_i}$ as the mutual eigenstates of $PH_{\mathrm{int}}P$  and $\Pi$.
We classify these dark eigenstates by their even and odd $j$-parity, corresponding to eigenvalues $1$ and $-1$ of $\Pi$.

The squeezed vacuum field induces transitions between the dark eigenstates, that result in the emission of a photon into the waveguide.
Since the jump operators are composed of linear combinations of the individual emitter operators $\sigma_i$ and $\sigma^\dagger_i$, they impose a strict $j$-parity selection rule: transitions are exclusively allowed between dark eigenstates possessing different $j$-parity.

To elucidate the transition mechanism, we first examine a minimal system of $N=4$ emitters with weak squeezing $r=0.01$ (Fig.~\ref{fig:DressedDiagram}).
For $N=4$, the dark-state subspace is 6 dimensional and spanned by the dark states $\ket{D_i}$ illustrated in Fig.~\ref{fig:DressedDiagram}(a).
In this case, the projected Hamiltonian $PH_{\mathrm{int}}P$ has three distinct energy levels. 
Its dark eigenstates are
$\ket{\psi_1} = \frac{1}{\sqrt{2}}\left( \ket{D_1} + \ket{D_6} \right), \,
\ket{\psi_6} = \frac{1}{\sqrt{2}}\left( \ket{D_1} -\ket{D_6} \right)$ and $\ket{\psi_i} = \ket{D_i}$ for $i\in \{2,3,4,5 \}
$.
The allowed transitions are between the odd $j$-parity states: $\ket{\psi_2},\ket{\psi_3},\ket{\psi_4}$, and the even $j$-parity states: $\ket{\psi_1},\ket{\psi_5},\ket{\psi_6}$ only.
These transitions are shown in Fig.~\ref{fig:DressedDiagram}(b), which also illustrates the three possible energy levels in this system, relative to the transition energy $\hbar \omega$ of the emitters.
The three distinct transition frequencies are plotted as blue and orange dashed lines in Fig.~\ref{fig:DressedDiagram}(c), corresponding to the color-coded arrows in Fig.~\ref{fig:DressedDiagram}(b).

The resulting incoherent power spectrum, normalized to its maximal value, is calculated using Eq.~(\ref{eq:power spectrum}) and plotted in Fig.~\ref{fig:DressedDiagram}(c), showing three distinct peaks that match the allowed transitions. The spectrum shape resembles the famous Mollow triplet \cite{mollow1969power}, corresponding to transitions between the dressed states of a single emitter driven by a laser. Here, however, the system comprises four emitters, dressed by the squeezed vacuum.  

\begin{figure*}
\includegraphics[width= \linewidth]{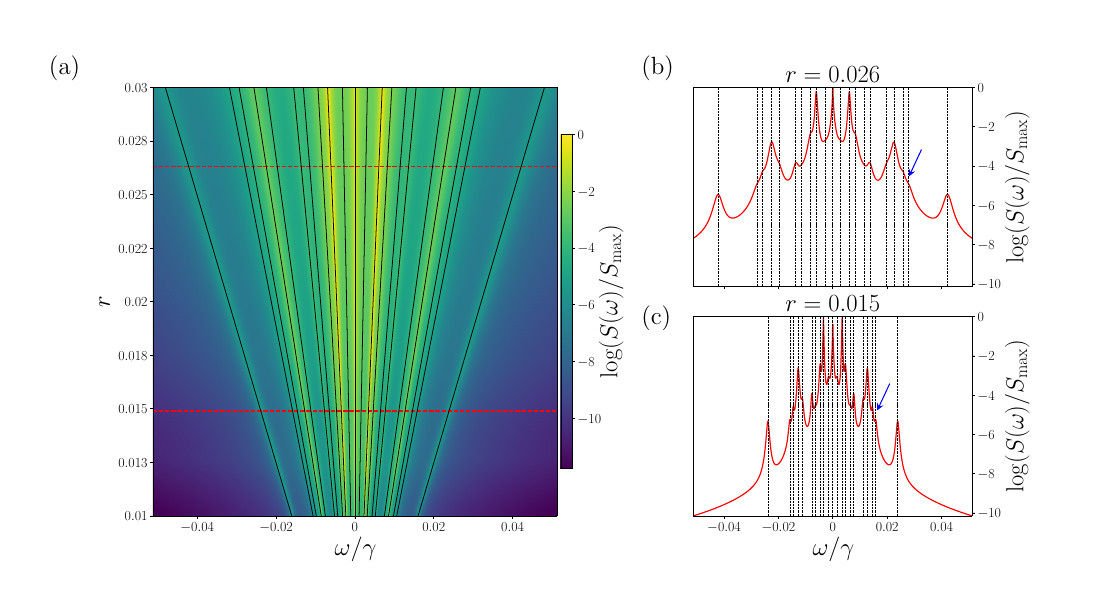}
\caption{\label{fig:6 emiiters ideal}
\textbf{Power spectrum for $N=6$ emitters across varying squeezing parameters in the ideal case ($\gamma_0=0$).} 
\textbf{(a)} Heatmap of the incoherent steady-state power spectrum in logarithmic scale ($\log_{10}$), normalized for each value of $r$ independently.
The solid black lines correspond to allowed transition frequencies between dark eigenstates of the projected Hamiltonian $PH_{\mathrm{int}}P$. 
They show excellent agreement with the power spectrum peaks for small squeezing parameter $r$.
For increasing values of $r$, the spectral features are broadened and spaced apart, making the peaks less distinct.
\textbf{(b)}, \textbf{(c)} Spectral line cross sections corresponding to the horizontal red lines in panel (a) for $r=0.026$ and $r=0.015$, respectively. Vertical dashed lines denote the allowed transition frequencies. 
The blue arrows in panel (c) are pointing at respective peaks of the power spectrum, demonstrating that increasing $r$ broadens the transition peaks and moves them farther apart.
}
\end{figure*}

The resemblance to the Mollow triplet spectral shape breaks down when we consider a larger system, yielding significantly richer dynamics, and resulting in more features in the power spectrum.
For example, we expand the system  to $N=6$ emitters.
In this case, the dark-state manifold becomes 20-dimensional, and there are $21$ distinct parity-allowed transition frequencies.
These frequencies are plotted as black lines in Fig.\ref{fig:6 emiiters ideal}(a), on top of the numerically calculated incoherent power spectrum heatmap. 
The incoherent power spectrum is independently normalized per $r$ value to its maximal value as a function of $\omega$, and is plotted in logarithmic ($\log_{10}$) scale.
The power spectrum for two of the $r$ values within the range are plotted in Fig.~\ref{fig:6 emiiters ideal}(b) and Fig.~\ref{fig:6 emiiters ideal}(c), with the respective $r$ values depicted as red horizontal lines in Fig.~\ref{fig:6 emiiters ideal}(a). The transition frequencies in Fig.~\ref{fig:6 emiiters ideal}(a) are plotted as gray dashed vertical lines in Fig.~\ref{fig:6 emiiters ideal}(b) and (c).
The results demonstrate complete alignment of the allowed transition frequencies with the peaks of the power spectrum for small values of $r$, even for a larger system.

Increasing the value of $r$ broadens the transition peaks and moves them farther apart. 
This trend increases the overlap of different features in the power spectrum, as can be seen by comparing Fig.~\ref{fig:6 emiiters ideal}(b) and Fig.~\ref{fig:6 emiiters ideal}(c). 
Some of the well defined peaks in Fig.~\ref{fig:6 emiiters ideal}(c) are barely distinct in Fig.~\ref{fig:6 emiiters ideal}(b), as can be seen, for instance, for the second- and third-rightmost peaks that are marked by blue arrows.

\section{\label{sec:Experimental considerations}Experimental considerations}

In any realistic waveguide quantum electrodynamics implementation, emitters inevitably exhibit some degree of coupling to non-guided electromagnetic modes~\cite{blais2021circuit}. 
This parasitic decay into free-space is characterized by the rate $\gamma_0$, breaking the perfect destructive interference that defines purely dark states and permits their potentially infinite lifetimes.
To quantify the impact of this nonideality on our spectroscopic protocol, we define the waveguide coupling efficiency as $\beta = \gamma / (\gamma + \gamma_0)$.

\begin{figure}
\includegraphics[width= \linewidth]{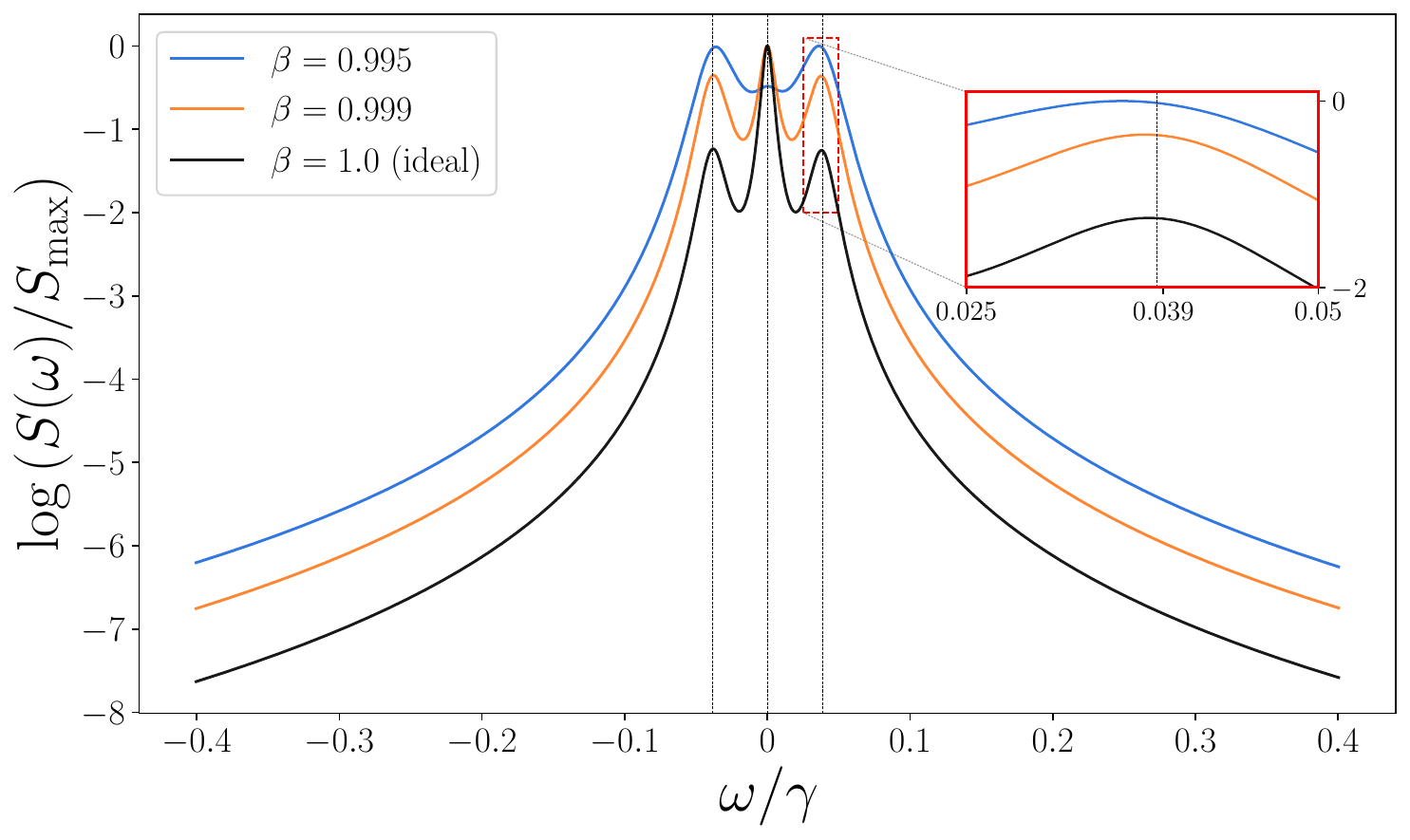}
\caption{\label{fig:4 emiiters nonideal}
\textbf{Incoherent power spectrum for a nonideal system.}
Incoherent power spectrum at steady-state in logarithmic scale ($\log_{10}$) are plotted for different values of waveguide coupling efficiency $\beta = \gamma / \left( \gamma + \gamma_0 \right)$.
$\beta=1$ refers to the ideal case where $\gamma_0=0$.
The plots are calculated with squeezing parameter $r=0.01$ and waveguide decay rate $\gamma = 1.0$.
Inset is a zoom-in around the peaks of the curves. 
It shows broadening and Fano-like shifts in the peak position relative to the transition frequencies.
}
\end{figure}

Fig.~\ref{fig:4 emiiters nonideal} illustrates the steady-state incoherent power spectrum of the right-moving waveguide mode for an array of $N=4$ emitters under varying coupling efficiencies $\beta$. 
The ideal case ($\beta=1$) exhibits sharp, symmetric peaks positioned exactly at the allowed transition frequencies.
However, as the coupling to free space increases ($\beta = 0.999$ and $\beta = 0.995$), two distinct spectral changes emerge.
First, the resonance peaks undergo significant broadening.
This is an expected consequence of the free-space decay introducing an additional decoherence channel, which fundamentally decreases the lifetime of the dark states.
Second, as highlighted in the inset of Fig.~\ref{fig:4 emiiters nonideal}, the resonance peaks exhibit Fano-like frequency shifts relative to the analytically calculated transition lines.

Despite these nonideal effects, the spectral signatures corresponding to the dark-manifold transitions remain highly distinguishable for realistic systems with high coupling efficiency.
A prime physical realization of this architecture is found in superconducting circuits, where artificial atoms (such as transmon qubits) coupled to 1D microwave transmission lines achieve coupling efficiencies of $\beta \approx 0.99 - 0.999$~\cite{shermet2023waveguidereview, mirhosseini2019cavity,hoi2011demonstration,astafiev2010resonance}. 
Furthermore, the required nonlinear medium can be directly realized using Josephson junction arrays.
By embedding thousands of Josephson junctions into a transmission line, the structure acts as a nonlinear metamaterial, forming a Josephson traveling-wave parametric amplifier~\cite{macklin2015near, zorin2016josephson,qiu2023broadband}, which can provide the distributed parametric gain and $\chi^{(2)}$ three-wave mixing~\cite{zorin2016josephson} described in our model, establishing the experimental feasibility of our proposed spectroscopic technique.

\section{\label{sec:Discussion}Discussion}

In this work, we demonstrated that introducing a $\chi^{(2)}$ optical nonlinearity into a WQED system provides a powerful mechanism for the direct spectroscopy of completely dark collective states.
By coherently pumping the nonlinear medium to create a traveling-wave parametric amplifier, we generate squeezed vacuum fields that mediate long-range interactions and break the permutation symmetry of the emitter array.
The pumps act as an activation switch: activating the drive couples the otherwise inaccessible dark states to the propagating waveguide modes for readout and control, while deactivating it completely decouples them, restoring their infinite lifetimes.

In the weak squeezing limit, the system dynamics are completely contained within the dark state manifold governed by an interaction Hamiltonian and collective jump operators (Eq.~(\ref{eq:SLH_dynamics})). 
We showed that these jump operators allow dark states to radiate into the waveguide exclusively through $j$-parity changing transitions.
Through numerical analysis of $N=4$ and $N=6$ emitter arrays, we confirmed that the steady-state incoherent power spectrum maps perfectly to the transition frequencies within the dark state manifold. 
Furthermore, our evaluation of nonideal systems indicates that this spectroscopic readout remains robust under realistic experimental conditions where finite free-space decay is present.

These results open several theoretical and experimental avenues.
The ability to control the long-range interactions via the squeezing parameter $r$ offers a direct knob for manipulating the dark-state manifold.
While this work focuses on spectroscopic readout, the underlying mechanism holds great potential for the deterministic preparation and coherent control of many-body entangled states.
Ultimately, our findings establish nonlinear WQED as a highly promising platform for advancing quantum memories and scalable quantum computational protocols reliant on robust, decoherence-free subspaces.

\newpage
\bibliography{Bibliography}

\end{document}